\documentstyle[thmsa]{article}
\usepackage{sw20lart}

\input tcilatex
\QQQ{Language}{
British English
}

\begin{document}

\section{Introduction}

Quantum mechanical probability amplitudes may be multiplied by an arbitrary
phase factor without affecting measured quantities which depend on them. We
have recently introduced generalized spin quantities whose derivation
depends upon probability amplitudes characterizing spin-projection
measurements $[1-3]$. In these derivations, the phase of these probability
amplitudes was arbitrarily chosen. In this paper, we investigate the effect
of changing the phase. We find that, as expected, when we change the phase
of one of the probability amplitudes, both the operators for the generalized
spin components and their eigenvectors change form. Focusing our attention
on the spin-$1/2$ case, we obtain alternative forms for the components of
the spin operator and for the spin states. We give the eigenvectors of these
alternative forms and explain how these can be easily computed for whatever
phase choice we make.

This paper is organized as follows. The general theory underlying the method
used to obtain the generalized spin quantities is presented in Section $2$.
Thus, the general implicit formulas for the $z$ component of the spin are
given in Subsection $2.1$. The method by which the generalized probability
amplitudes are derived is sketched in Subsection $2.2$. This theory is then
illustrated in Subsection $2.3$ where, using one phase choice, the explicit
formulas for the probability amplitudes, the three spin components and their
eigenvalues are given. In Section $3$, we change the phase and thus obtain
different probability amplitudes. These are presented in Subsection $3.1$.
The new $z$ component of spin and its eigenvalues resulting from this change
in probability amplitudes are given in Subsection $3.2$, while the $x$ and $%
y $ components of spin and their eigenvectors are presented in Subsections $%
3.3 $ and $3.4$. In Section $4$, we relate the new results to standard ones
by giving the limit in which the new results reduce to the standard ones.
The Discussion and Conclusion in Section $5$ close the paper.

\section{General Formulas}

\subsection{z Component of the Spin Operator}

In this paper, we are concerned with the case of spin $1/2$. We measure spin
in units of $\hbar /2.$ Let $\widehat{{\bf a}}$ be an arbitrary vector whose
polar angles are $(\theta ^{\prime \prime },\varphi ^{\prime \prime })$, and
let $\widehat{{\bf c}}$ be the vector with polar angles $(\theta ^{\prime
},\varphi ^{\prime }).$ We wish to measure the spin projection with respect
to the vector $\widehat{{\bf c}}$ after having measured it with respect to
the vector $\widehat{{\bf a}}$. Thus, the spin projection is initially known
to be either up or down with respect to the vector $\widehat{{\bf a}}$. We
denote by $\psi $ the probability amplitudes for the spin projection
measurement $.$ Thus, if the spin projection is initially up with respect to 
$\widehat{{\bf a}}$, the probability amplitude for finding it up with
respect to $\widehat{{\bf c}}$ upon measurement is $\psi ((+\frac 12)^{(%
\widehat{{\bf a}})};(+\frac 12)^{(\widehat{{\bf c}})}).$ There are three
other probability amplitudes: these are $\psi ((+\frac 12)^{(\widehat{{\bf a}%
})};(-\frac 12)^{(\widehat{{\bf c}})})$, $\psi ((-\frac 12)^{(\widehat{{\bf a%
}})};(+\frac 12)^{(\widehat{{\bf c}})})$ and $\psi ((-\frac 12)^{(\widehat{%
{\bf a}})};(-\frac 12)^{(\widehat{{\bf c}})})$:$.$ their interpretation is
obvious.

The probability ampplitudes $\psi $ can be expanded in terms of the
probability amplitudes $\chi $ and $\phi $. The probability amplitudes $\chi 
$ describe spin-projection measurements from the quantization direction $%
\widehat{{\bf a}}$ to the quantization direction $\widehat{{\bf b}}$, where
the vector $\widehat{{\bf b}}$ has the polar angles $(\theta ,\varphi ).$
The expansions are 
\begin{eqnarray}
\psi ((+\frac 12)^{(\widehat{{\bf a}})};(+\frac 12)^{(\widehat{{\bf c}})})
&=&\chi ((+\frac 12)^{(\widehat{{\bf a}})};(+\frac 12)^{(\widehat{{\bf b}}%
)})\phi ((+\frac 12)^{(\widehat{{\bf b}})};(+\frac 12)^{(\widehat{{\bf c}})})
\nonumber  \label{bone1} \\
&&\ +\chi ((+\frac 12)^{(\widehat{{\bf a}})};(-\frac 12)^{(\widehat{{\bf b}}%
)})\phi ((-\frac 12)^{(\widehat{{\bf b}})};(+\frac 12)^{(\widehat{{\bf c}}%
)}),  \label{aone1}
\end{eqnarray}

\begin{eqnarray}
\psi ((+\frac 12)^{(\widehat{{\bf a}})};(-\frac 12)^{(\widehat{{\bf c}})})
&=&\chi ((+\frac 12)^{(\widehat{{\bf a}})};(+\frac 12)^{(\widehat{{\bf b}}%
)})\phi ((+\frac 12)^{(\widehat{{\bf b}})};(-\frac 12)^{(\widehat{{\bf c}})})
\nonumber \\
&&\ \ +\chi ((+\frac 12)^{(\widehat{{\bf a}})};(-\frac 12)^{(\widehat{{\bf b}%
})}))\phi ((-\frac 12)^{(\widehat{{\bf b}})};(-\frac 12)^{(\widehat{{\bf c}}%
)}),  \label{zone1}
\end{eqnarray}
\begin{eqnarray}
\psi ((-\frac 12)^{(\widehat{{\bf a}})};(+\frac 12)^{(\widehat{{\bf c}})})
&=&\chi ((-\frac 12)^{(\widehat{{\bf a}})};(+\frac 12)^{(\widehat{{\bf b}}%
)})\phi ((+\frac 12)^{(\widehat{{\bf b}})};(+\frac 12)^{(\widehat{{\bf c}})})
\nonumber \\
&&\ \ +\chi ((-\frac 12)^{(\widehat{{\bf a}})};(-\frac 12)^{(\widehat{{\bf b}%
})})\phi ((-\frac 12)^{(\widehat{{\bf b}})};(+\frac 12)^{(\widehat{{\bf c}}%
)})  \label{cone1}
\end{eqnarray}
and 
\begin{eqnarray}
\psi ((-\frac 12)^{(\widehat{{\bf a}})};(-\frac 12)^{(\widehat{{\bf c}})})
&=&\chi ((-\frac 12)^{(\widehat{{\bf a}})};(+\frac 12)^{(\widehat{{\bf b}}%
)})\phi ((+\frac 12)^{(\widehat{{\bf b}})};(-\frac 12)^{(\widehat{{\bf c}})})
\nonumber \\
&&\ \ +\chi ((-\frac 12)^{(\widehat{{\bf a}})};(-\frac 12)^{(\widehat{{\bf b}%
})})\phi ((-\frac 12)^{(\widehat{{\bf b}})};(-\frac 12)^{(\widehat{{\bf c}}%
)}).  \label{done1}
\end{eqnarray}

Let the quantity $R({\bf \sigma }\cdot \widehat{{\bf c}})$ have the value $%
r_1$ when the spin projection is up (quantum number $m_1$) with respect to
the vector $\widehat{{\bf c}}$, and $r_{2\;}$when it is down (quantum number 
$m_2$). Thus, the possible values of $R$ are $r_n$ $(n=1,2)$. Suppose that
the initial state corresponds to the quantum number $m_i$ with respect to $%
\widehat{{\bf a}}$. As the probability amplitude for obtaining $r_n$ is $%
\psi (m_i^{(\widehat{{\bf a}})};m_n^{(\widehat{{\bf c}})})$, the expectation
value of $R$ is

\begin{equation}
\left\langle R\right\rangle =\sum_{n=1}^2\left| \psi (m_i^{(\widehat{{\bf a}}%
)};m_n^{(\widehat{{\bf c}})})\right| ^2r_n.  \label{atwo}
\end{equation}

Since the expansions for $\psi ^{*}(m_i^{(\widehat{{\bf a}})};m_n^{(\widehat{%
{\bf c}})})$ and $\psi (m_i^{(\widehat{{\bf a}})};m_n^{(\widehat{{\bf c}})})$
are

\begin{equation}
\psi ^{*}(m_i^{(\widehat{{\bf a}})};m_n^{(\widehat{{\bf c}}%
)})=\sum_{j=1}^2\chi ^{*}(m_i^{(\widehat{{\bf a}})};m_j^{(\widehat{{\bf b}}%
)})\phi ^{*}(m_j^{(\widehat{{\bf b}})};m_n^{(\widehat{{\bf c}})})
\label{athree}
\end{equation}
and

\begin{equation}
\psi (m_i^{(\widehat{{\bf a}})};m_n^{(\widehat{{\bf c}})})=\sum_{j^{\prime
}=1}^2\chi (m_i^{(\widehat{{\bf a}})};m_{j^{\prime }}^{(\widehat{{\bf b}}%
)})\phi (m_{j^{\prime }}^{(\widehat{{\bf b}})};m_n^{(\widehat{{\bf c}})}),
\label{afour}
\end{equation}
it follows that

\begin{eqnarray}
\left\langle R\right\rangle &=&\sum_j\sum_{j^{\prime }}\chi ^{*}(m_i^{(%
\widehat{{\bf a}})};m_j^{(\widehat{{\bf b}})})R_{jj^{\prime }}\chi (m_i^{(%
\widehat{{\bf a}})};m_{j^{\prime }}^{(\widehat{{\bf b}})})  \nonumber \\
\ &=&[\psi (m_i^{(\widehat{{\bf a}})};m_n^{(\widehat{{\bf c}})})]^{\dagger
}[R][\psi (m_i^{(\widehat{{\bf a}})};m_n^{(\widehat{{\bf c}})})]
\label{afive}
\end{eqnarray}
where [$1$] 
\begin{equation}
\lbrack \psi (m_i^{(\widehat{{\bf a}})};m_n^{(\widehat{{\bf c}})})]=\left( 
\begin{array}{c}
\chi (m_i^{(\widehat{{\bf a}})},(+\frac 12)^{(\widehat{{\bf b}})}) \\ 
\chi (m_i^{(\widehat{{\bf a}})},(-\frac 12)^{(\widehat{{\bf b}})})
\end{array}
\right)  \label{seven}
\end{equation}
and 
\begin{equation}
\left[ R\right] =\left( 
\begin{array}{cc}
R_{11} & R_{12} \\ 
R_{21} & R_{22}
\end{array}
\right) ,  \label{aseven}
\end{equation}
with the elements of $[R]$ being given by

\begin{equation}
R_{jj^{\prime }}=\sum_{n=1}^2\phi ^{*}(m_j^{(\widehat{{\bf b}})};m_n^{(%
\widehat{{\bf c}})})\phi (m_{j^{\prime }}^{(\widehat{{\bf b}})};m_n^{(%
\widehat{{\bf c}})})r_n.  \label{aeight}
\end{equation}

Written out explicitly, the elements of $\;\left[ R\right] \;$are

\begin{equation}
R_{11}=\left| \phi ((+\frac 12)^{(\widehat{{\bf b}})};(+\frac 12)^{(\widehat{%
{\bf c}})})\right| ^2r_1+\left| \phi ((+\frac 12)^{(\widehat{{\bf b}}%
)};(-\frac 12)^{(\widehat{{\bf c}})})\right| ^2r_2,  \label{tw2}
\end{equation}

\begin{eqnarray}
R_{12} &=&\phi ^{*}((+\frac 12)^{(\widehat{{\bf b}})};(+\frac 12)^{(\widehat{%
{\bf c}})})\phi ((-\frac 12)^{(\widehat{{\bf b}})};(+\frac 12)^{(\widehat{%
{\bf c}})})r_1  \nonumber \\
&&+\phi ^{*}((+\frac 12)^{(\widehat{{\bf b}})};(-\frac 12)^{(\widehat{{\bf c}%
})})\phi ((-\frac 12)^{(\widehat{{\bf b}})};(-\frac 12)^{(\widehat{{\bf c}}%
)})r_2,  \label{th3}
\end{eqnarray}

\begin{eqnarray}
R_{21} &=&\phi ^{*}((-\frac 12)^{(\widehat{{\bf b}})};(+\frac 12)^{(\widehat{%
{\bf c}})})\phi ((+\frac 12)^{(\widehat{{\bf b}})};(+\frac 12)^{(\widehat{%
{\bf c}})})r_1  \nonumber \\
&&+\phi ^{*}((-\frac 12)^{(\widehat{{\bf b}})};(-\frac 12)^{(\widehat{{\bf c}%
})})\phi ((+\frac 12)^{(\widehat{{\bf b}})};(-\frac 12)^{(\widehat{{\bf c}}%
)})r_2  \label{fo4}
\end{eqnarray}
and

\begin{equation}
R_{22}=\left| \phi ((-\frac 12)^{(\widehat{{\bf b}})};(+\frac 12)^{(\widehat{%
{\bf c}})})\right| ^2r_1+\left| \phi ((-\frac 12)^{(\widehat{{\bf b}}%
)};(-\frac 12)^{(\widehat{{\bf c}})})\right| ^2r_2.  \label{fi5}
\end{equation}

When $R={\bf \sigma }\cdot \widehat{{\bf c}}$, then it is the component of
the spin in the direction $\widehat{{\bf c}}$. Hence, $\left[ R\right] $ is
the matrix form of the operator for the component of the spin along the axis
defined by $\widehat{{\bf c}}$. In its most generalized form, we shall
denote this operator by $[\sigma _{\widehat{{\bf c}}}]$.

When $R=\sigma ^2$, then $[R]$ is the matrix form of the square of the total
spin.

The precise form of both the vector state Eqn. (\ref{seven}) and the
operator Eqns. (\ref{tw2}) - (\ref{fi5}) is evidently a function of the
phase adopted for the probability amplitudes $\psi $, $\chi $ and $\phi $.
Since the phase does not change the probabilities, we see from Eqns. (\ref
{tw2}) and (\ref{fi5}) that the diagonal elements remain the same no matter
what phase we choose. But the form of the off-diagonal elements is a
function of the choice of phase. In order to make these observations
concrete, we need to obtain explicit expressions for the probability
amplitudes.

\subsection{Probability Amplitudes}

Following the method we introduced in Ref. $1$ and used in Refs. $2$ and $3$%
, we obtain the probability amplitudes in the following manner. Consider the
vector $\widehat{{\bf a}}$ (whose polar angles are $(\theta ^{\prime \prime
},\varphi ^{\prime \prime })).$ The dot product of this vector with the
standard spin operator

\begin{equation}
\left[ {\bf \sigma }\right] =\widehat{{\bf i}}\left( 
\begin{array}{cc}
0 & 1 \\ 
1 & 0
\end{array}
\right) +\widehat{{\bf j}}\left( 
\begin{array}{cc}
0 & -i \\ 
i & 0
\end{array}
\right) +\widehat{{\bf k}}\left( 
\begin{array}{cc}
1 & 0 \\ 
0 & -1
\end{array}
\right)  \label{eight}
\end{equation}
is

\begin{equation}
\lbrack {\bf \sigma \cdot }\widehat{{\bf a}}]=\left( 
\begin{array}{cc}
\cos \theta ^{\prime \prime } & \sin \theta ^{\prime \prime }e^{-i\varphi
^{\prime \prime }} \\ 
\sin \theta e^{i\varphi ^{\prime \prime }} & -\cos \theta ^{\prime \prime }
\end{array}
\right)  \label{nine}
\end{equation}
This operator is the spin projection in the direction $\widehat{{\bf a}}$.
As it is the generalized form of the operator for $z$ component of the spin,
we shall call it the ''standard generalized $z$ component'' of the spin.
Corresponding to it are the ''standard generalized eigenvectors''.
Similarly, the $x$ and $y$ components that go with this operator, as well as
their eigenvectors are called ''standard generalized forms''. The reason is
that they appear to be the most generalized forms for these quantities
appearing in the literature. However, as we have already shown in Refs. $1-3$%
, these are in fact specialized forms of even more generalized quantities.

The elements of a matrix eigenvector are probability amplitudes. As proven
in Ref. $1$, these probability amplitudes are not just constants but have a
structure. Thus, in this case the eigenvectors for the eigenvalues $+1$ and $%
-1$ have the respective forms

\begin{equation}
\lbrack \xi _{+}]=\left( 
\begin{array}{c}
\chi ((+\frac 12)^{(\widehat{{\bf a}})};(+\frac 12)^{(\widehat{{\bf f}})})
\\ 
\chi ((+\frac 12)^{(\widehat{{\bf a}})};(-\frac 12)^{(\widehat{{\bf f}})})
\end{array}
\right)  \label{ten}
\end{equation}
and

\begin{equation}
\lbrack \xi _{-}]=\left( 
\begin{array}{c}
\chi ((-\frac 12)^{(\widehat{{\bf a}})};(+\frac 12)^{(\widehat{{\bf f}})})
\\ 
\chi ((-\frac 12)^{(\widehat{{\bf a}})};(-\frac 12)^{(\widehat{{\bf f}})})
\end{array}
\right)  \label{el11}
\end{equation}
where $\widehat{{\bf f}}$ is an unknown direction vector.

On the other hand, for $\widehat{{\bf c}}$ (whose polar angles are $(\theta
^{\prime },\varphi ^{\prime }))$, we have 
\begin{equation}
\lbrack {\bf \sigma \cdot }\widehat{{\bf c}}]=\left( 
\begin{array}{cc}
\cos \theta ^{\prime } & \sin \theta ^{\prime }e^{-i\varphi ^{\prime }} \\ 
\sin \theta ^{\prime }e^{i\varphi ^{\prime }} & -\cos \theta ^{\prime }
\end{array}
\right)  \label{tw12}
\end{equation}

The eigenvectors of this matrix for the eigenvalues $+1$ and $-1$ are
respectively 
\begin{equation}
\lbrack \xi _{+}]=\left( 
\begin{array}{c}
\chi ((+\frac 12)^{(\widehat{{\bf c}})};(+\frac 12)^{(\widehat{{\bf f}})})
\\ 
\chi ((+\frac 12)^{(\widehat{{\bf c}})};(-\frac 12)^{(\widehat{{\bf f}})})
\end{array}
\right)  \label{th13}
\end{equation}
and

\begin{equation}
\lbrack \xi _{-}]=\left( 
\begin{array}{c}
\chi ((-\frac 12)^{(\widehat{{\bf c}})};(+\frac 12)^{(\widehat{{\bf f}})})
\\ 
\chi ((-\frac 12)^{(\widehat{{\bf c}})};(-\frac 12)^{(\widehat{{\bf f}})})
\end{array}
\right)  \label{fo14}
\end{equation}

We now use the Land\'e expansion [$4-7$] for probability amplitudes to
obtain the generalized probability amplitudes $\chi (m_i^{\widehat{{\bf a}}%
};m_n^{\widehat{{\bf c}}})$, where $i,n=1,2.$ This expansion concerns three
observables $A$, $B$ and $C$ which belong to the same quantum system. These
observables are characterized by the states corresponding to the eigenvalues 
$A_i,$ $B_j$ and $C_k$ respectively. The probability amplitudes connecting
states of these observables are inter-related in the following way:

\begin{equation}
\psi (A_i;C_n)=\dsum\limits_j\xi (A_i;B_j)\phi (B_j;C_n)  \label{fi15}
\end{equation}
where $\psi ,$ $\xi $ and $\phi $ are all probability amplitudes and $j$
runs over all the states corresponding to $B$. In this case, all three
observables are spin projections; thus, $A\rightarrow \widehat{{\bf a}}$, $%
B\rightarrow \widehat{{\bf f}}$ and $C\rightarrow \widehat{{\bf c}}$. Owing
to the fact that all three probability amplitudes refer to spin projection
measurements, they have the same structure; for this reason, we shall for
the moment collectively denote them by $\chi $. Using the Hermiticity
property of the probability amplitudes [4] 
\begin{equation}
\chi (A_i;C_n)=\chi ^{*}(C_n;A_i),  \label{si16}
\end{equation}
we are able to apply the Land\'e expansion, Eqn. (\ref{fi15}), in order to
eliminate the unknown vector $\widehat{{\bf f}}$ and so obtain the
probability amplitudes $\chi (m_i^{\widehat{{\bf a}}};m_n^{\widehat{{\bf c}}%
}).$

\subsection{Spin States}

The spin states, Eqn. (\ref{seven}), are seen to have a form that depends on
the expressions for the probability amplitudes. Their general form is
identical to the general form of the eigenvectors of spin operators. A
change in the phase of the probability amplitudes alters the form of the
spin states. Some of these states are special in the sense that they are
eigenvectors of spin operators, but this specialness consists in the
arguments which the expressions for the probability amplitudes have.

\subsection{Spin Quantities For Old Choice of Phase}

As observed above, the form of the generalized probability amplitudes
depends on the phase choice we make when we determine the vectors Eqns. (\ref
{ten}), (\ref{el11}), (\ref{th13}) and (\ref{fo14}). Our phase choice in
Refs. $1$- $3$ led to

\begin{equation}
\lbrack \xi _{+}]=\left( 
\begin{array}{c}
\cos \theta ^{\prime \prime }/2 \\ 
e^{i\varphi ^{\prime \prime }}\sin \theta ^{\prime \prime }/2
\end{array}
\right)  \label{se17}
\end{equation}
and 
\begin{equation}
\lbrack \xi _{-}]=\left( 
\begin{array}{c}
\sin \theta ^{\prime \prime }/2 \\ 
-e^{i\varphi ^{\prime \prime }}\cos \theta ^{\prime \prime }/2
\end{array}
\right)  \label{ei18}
\end{equation}
with identical expressions for $[\xi _{+}]$ and $[\xi _{-}],$but with
argument transformations $\theta ^{\prime \prime }\rightarrow \theta
^{\prime }$ and $\varphi ^{\prime \prime }\rightarrow \varphi ^{\prime }$.
The use of the Land\'e expansion Eqn. (\ref{fi15}) together with the
condition Eqn. (\ref{si16}) led to the generalized probability amplitudes 
\begin{equation}
\chi ((+\frac 12)^{(\widehat{{\bf a}})},(+\frac 12)^{(\widehat{{\bf c}}%
)})=\cos \theta ^{\prime \prime }/2\cos \theta ^{\prime }/2+e^{i(\varphi
^{\prime \prime }-\varphi ^{\prime })}\sin \theta ^{\prime \prime }/2\sin
\theta ^{\prime }/2  \label{ni19}
\end{equation}
\begin{equation}
\chi ((+\frac 12)^{(\widehat{{\bf a}})},(-\frac 12)^{(\widehat{{\bf c}}%
)})=\cos \theta ^{\prime \prime }/2\sin \theta ^{\prime }/2-e^{i(\varphi
^{\prime \prime }-\varphi ^{\prime })}\sin \theta ^{\prime \prime }/2\cos
\theta ^{\prime }/2  \label{tw20}
\end{equation}

\begin{equation}
\chi ((-\frac 12)^{(\widehat{{\bf a}})},(+\frac 12)^{(\widehat{{\bf c}}%
)})=\sin \theta ^{\prime \prime }/2\cos \theta ^{\prime }/2-e^{i(\varphi
^{\prime \prime }-\varphi ^{\prime })}\cos \theta ^{\prime \prime }/2\sin
\theta ^{\prime }/2  \label{tw21}
\end{equation}

\begin{equation}
\chi ((-\frac 12)^{(\widehat{{\bf a}})},(-\frac 12)^{(\widehat{{\bf c}}%
)})=\sin \theta ^{\prime \prime }/2\sin \theta ^{\prime }/2+e^{i(\varphi
^{\prime \prime }-\varphi ^{\prime })}\cos \theta ^{\prime \prime }/2\cos
\theta ^{\prime }/2  \label{tw22}
\end{equation}

Using these forms for the probability amplitudes in the general formulas
Eqns. (\ref{tw2}) - (\ref{fi5}), and remembering that the $\phi $'s have the
same form as the $\chi $'s, we are able to obtain explicit formulas for the
elements of $[\sigma _{\widehat{{\bf c}}}]$. We have to remember that as
well as changing from the $\chi $'s to the $\phi $'s, we have to change the
arguments to $\phi (m_i^{\widehat{{\bf b}}};m_n^{\widehat{{\bf c}}}),$ where
the polar angles of $\widehat{{\bf b}}$ are $(\theta ,\varphi ).$ We find
that the expressions for the elements of the operator $[\sigma _{\widehat{%
{\bf c}}}]$ are:

\begin{equation}
(\sigma _{\widehat{{\bf c}}})_{11}=\cos \theta \cos \theta ^{\prime }+\sin
\theta \sin \theta ^{\prime }\cos (\varphi -\varphi ^{\prime })  \label{tw23}
\end{equation}

\begin{equation}
(\sigma _{\widehat{{\bf c}}})_{12}=\sin \theta \cos \theta ^{\prime }-\sin
\theta \cos \theta ^{\prime }-\sin \theta ^{\prime }[\cos \theta \cos
(\varphi -\varphi ^{\prime })+i\sin (\varphi -\varphi ^{\prime })]
\label{tw24}
\end{equation}

\begin{equation}
(\sigma _{\widehat{{\bf c}}})_{21}=\sin \theta \cos \theta ^{\prime }-\sin
\theta \cos \theta ^{\prime }-\sin \theta ^{\prime }[\cos \theta \cos
(\varphi -\varphi ^{\prime })-i\sin (\varphi -\varphi ^{\prime })]
\label{tw25}
\end{equation}
and

\begin{equation}
(\sigma _{\widehat{{\bf c}}})_{11}=-\cos \theta \cos \theta ^{\prime }-\sin
\theta \sin \theta ^{\prime }\cos (\varphi -\varphi ^{\prime })  \label{tw26}
\end{equation}

In order to obtain the eigenvectors of $[\sigma _{\widehat{{\bf c}}}]$, we
use the fact that while the probability amplitudes $\phi (m_i^{\widehat{{\bf %
b}}};m_n^{\widehat{{\bf c}}})$ are needed obtain the elements of $[\sigma _{%
\widehat{{\bf c}}}]$, the elements of the eigenvectors are of the form $\chi
(m_i^{\widehat{{\bf c}}};m_n^{\widehat{{\bf b}}})$, as deduced in Ref. $1$.
Thus, the eigenvectors of this operator are

\begin{equation}
\lbrack \xi _{\widehat{{\bf c}}}^{(+)}]=\left( 
\begin{array}{c}
\chi ((+\frac 12)^{(\widehat{{\bf c}})},(+\frac 12)^{(\widehat{{\bf b}})})
\\ 
\chi ((+\frac 12)^{(\widehat{{\bf c}})},(-\frac 12)^{(\widehat{{\bf b}})})
\end{array}
\right) =\left( 
\begin{array}{c}
\cos \theta ^{\prime }/2\cos \theta /2+e^{i(\varphi ^{\prime }-\varphi
)}\sin \theta ^{\prime }/2\sin \theta /2 \\ 
\cos \theta ^{\prime }/2\sin \theta /2-e^{i(\varphi ^{\prime }-\varphi
)}\sin \theta ^{\prime }/2\cos \theta /2
\end{array}
\right)  \label{tw27}
\end{equation}
for eigenvalue $+1$ and 
\begin{equation}
\lbrack \xi _{\widehat{{\bf c}}}^{(-)}]=\left( 
\begin{array}{c}
\chi ((-\frac 12)^{(\widehat{{\bf c}})},(+\frac 12)^{(\widehat{{\bf b}})})
\\ 
\chi ((-\frac 12)^{(\widehat{{\bf c}})},(-\frac 12)^{(\widehat{{\bf b}})})
\end{array}
\right) =\left( 
\begin{array}{c}
\sin \theta ^{\prime }/2\cos \theta /2-e^{i(\varphi ^{\prime }-\varphi
)}\cos \theta ^{\prime }/2\sin \theta /2 \\ 
\sin \theta ^{\prime }/2\sin \theta /2+e^{i(\varphi ^{\prime }-\varphi
)}\cos \theta ^{\prime }/2\cos \theta /2
\end{array}
\right)  \label{tw28}
\end{equation}
for eigenvalue $-1$.

The operator $[\sigma _{\widehat{{\bf c}}}]$ is a generalized form of the $z$
component of the spin. As we can see, it is more generalized than the
''standard generalized forms'' discussed earlier. Corresponding to $[\sigma
_{\widehat{{\bf c}}}]$ are operators which we may formally obtain through
the generalized ladder operators. The generalized ladder operators are
obtainable from their actions on the eigenvectors $[\xi _{\widehat{{\bf c}}%
}^{(+)}]$ and $[\xi _{\widehat{{\bf c}}}^{(-)}]$ of $[\sigma _{\widehat{{\bf %
c}}}].$ Then we use the definitions of the ladder operators in terms of the $%
x$ and $y$ components of the spin operator to obtain these quantities.
However, a shorter method was introduced in Refs. $2$ and $3$.

The operator $[\sigma _x]$ may be obtained for this case from the operator $%
[\sigma _{\widehat{{\bf c}}}]$ by setting $\theta ^{\prime }\rightarrow
\theta ^{\prime }-\pi /2$ and leaving $\varphi ^{\prime }$ unchanged [$2,3$%
]. The same transformation gives the eigenvectors of $[\sigma _x]$ from
those of $[\sigma _{\widehat{{\bf c}}}].$

The elements of $[\sigma _x]$ are found to be [$1,2$]

\begin{equation}
(\sigma _x)_{11}=-\sin \theta \cos \theta ^{\prime }\cos (\varphi ^{\prime
}-\varphi )+\sin \theta ^{\prime }\cos \theta  \label{th30}
\end{equation}

\begin{equation}
(\sigma _x)_{12}=\cos \theta \cos \theta ^{\prime }\cos (\varphi ^{\prime
}-\varphi )+\sin \theta \sin \theta ^{\prime }-i\cos \theta ^{\prime }\sin
(\varphi ^{\prime }-\varphi )  \label{th31}
\end{equation}

\begin{equation}
(\sigma _x)_{21}=\cos \theta \cos \theta ^{\prime }\cos (\varphi ^{\prime
}-\varphi )+\sin \theta \sin \theta ^{\prime }+i\cos \theta ^{\prime }\sin
(\varphi ^{\prime }-\varphi )  \label{th32}
\end{equation}
and

\begin{equation}
(\sigma _x)_{22}=\sin \theta \cos \theta ^{\prime }\cos (\varphi ^{\prime
}-\varphi )-\sin \theta ^{\prime }\cos \theta  \label{th33}
\end{equation}
The eigenvectors of $[\sigma _x]$ are [$2$]

\begin{equation}
\lbrack \xi _x^{(+)}]=\frac 1{\sqrt{2}}\left( 
\begin{array}{c}
(\sin \frac{\theta ^{\prime }}2+\cos \frac{\theta ^{\prime }}2)\cos \frac
\theta 2+e^{i(\varphi ^{\prime }-\varphi )}(\sin \frac{\theta ^{\prime }}%
2-\cos \frac{\theta ^{\prime }}2)\sin \frac \theta 2 \\ 
(\sin \frac{\theta ^{\prime }}2+\cos \frac{\theta ^{\prime }}2)\sin \frac
\theta 2-e^{i(\varphi ^{\prime }-\varphi )}(\sin \frac{\theta ^{\prime }}%
2-\cos \frac{\theta ^{\prime }}2)\cos \frac \theta 2
\end{array}
\right)  \label{th34}
\end{equation}
and 
\begin{equation}
\lbrack \xi _x^{(-)}]=\frac 1{\sqrt{2}}\left( 
\begin{array}{c}
(\sin \frac{\theta ^{\prime }}2-\cos \frac{\theta ^{\prime }}2)\cos \frac
\theta 2-e^{i(\varphi ^{\prime }-\varphi )}(\sin \frac{\theta ^{\prime }}%
2+\cos \frac{\theta ^{\prime }}2)\sin \frac \theta 2 \\ 
(\sin \frac{\theta ^{\prime }}2-\cos \frac{\theta ^{\prime }}2)\sin \frac
\theta 2+e^{i(\varphi ^{\prime }-\varphi )}(\sin \frac{\theta ^{\prime }}%
2+\cos \frac{\theta ^{\prime }}2)\cos \frac \theta 2
\end{array}
\right)  \label{th35}
\end{equation}
for the eigenvalues $+1$ and $-1$ respectively.

To obtain the operator $[\sigma _y]$ from $[\sigma _{\widehat{{\bf c}}}],$
we transform the arguments thus: $\theta ^{\prime }\rightarrow \pi /2$, $%
\varphi ^{\prime }\rightarrow \varphi ^{\prime }-\pi /2.$ We get the
eigenvectors of $[\sigma _y]$ by applying the same transformation to those
of $[\sigma _{\widehat{{\bf c}}}].$ The elements of $[\sigma _y]$ are [$1,2$]

\begin{equation}
(\sigma _y)_{11}=\sin \theta \sin (\varphi ^{\prime }-\varphi )  \label{th36}
\end{equation}

\begin{equation}
(\sigma _y)_{12}=-i\cos (\varphi ^{\prime }-\varphi )-\cos \theta \sin
(\varphi ^{\prime }-\varphi )  \label{th37}
\end{equation}

\begin{equation}
(\sigma _y)_{21}=i\cos (\varphi ^{\prime }-\varphi )-\cos \theta \sin
(\varphi ^{\prime }-\varphi )  \label{th38}
\end{equation}
and

\begin{equation}
(\sigma _y)_{22}=-\sin \theta \sin (\varphi ^{\prime }-\varphi )
\label{th39}
\end{equation}

The eigenvectors are [$2$]

\begin{equation}
\lbrack \chi _y^{(+)}]=\frac 1{\sqrt{2}}\left( 
\begin{array}{c}
\cos \frac \theta 2-ie^{i(\varphi ^{\prime }-\varphi )}\sin \frac \theta 2
\\ 
\sin \frac \theta 2+ie^{i(\varphi ^{\prime }-\varphi )}\cos \frac \theta 2
\end{array}
\right)  \label{fo40}
\end{equation}
and

\begin{equation}
\lbrack \chi _y^{(-)}]=\frac 1{\sqrt{2}}\left( 
\begin{array}{c}
\cos \frac \theta 2+ie^{i(\varphi ^{\prime }-\varphi )}\sin \frac \theta 2
\\ 
\sin \frac \theta 2-ie^{i(\varphi ^{\prime }-\varphi )}\cos \frac \theta 2
\end{array}
\right)  \label{fo41}
\end{equation}
corresponding to the eigenvalues $+1$ and $-1$ respectively.

We emphasize that the spin states have the same forms as the eigenvectors of
the $z$ component of spin. The two kinds of quantities differ only in the
arguments.

\section{New Choice of Phase}

\subsection{Probability Amplitudes}

We now choose a different phase for the probability amplitudes in the
vectors Eqns. (\ref{se17}) and (\ref{ei18}). This will lead to different
forms for the generalized probability amplitudes Eqns. (\ref{ni19}) and (\ref
{tw22}). Many alternatives are possible, but one alternative will suffice to
illustrate what happens to the spin quantities when we make this change of
phase. We effect this change in phase by taking for the solutions of 
\begin{equation}
\lbrack {\bf \sigma \cdot }\widehat{{\bf a}}]=\left( 
\begin{array}{cc}
\cos \theta ^{\prime \prime } & \sin \theta ^{\prime \prime }e^{-i\varphi
^{\prime \prime }} \\ 
\sin \theta ^{\prime \prime }e^{i\varphi ^{\prime \prime }} & -\cos \theta
^{\prime \prime }
\end{array}
\right)  \label{fo41d}
\end{equation}
the vectors 
\begin{equation}
\lbrack \xi _{-}]=\left( 
\begin{array}{c}
\sin \theta ^{\prime \prime }/2 \\ 
-e^{i\varphi ^{\prime \prime }}\cos \theta ^{\prime \prime }/2
\end{array}
\right)  \label{fo41z}
\end{equation}
and

\begin{equation}
\lbrack \xi _{+}]=\left( 
\begin{array}{c}
e^{-i\varphi ^{\prime \prime }}\cos \theta ^{\prime \prime }/2 \\ 
\sin \theta ^{\prime \prime }/2
\end{array}
\right)  \label{fo42}
\end{equation}
in place of Eqn. (\ref{se17}) and Eqn. (\ref{ei18}). Thus only Eqn. (\ref
{se17}) is changed; Eqn. (\ref{ei18}) remains unchanged. We do the same
thing when computing the eigenvectors of $[{\bf \sigma \cdot }\widehat{{\bf c%
}}].$ Using the Land\'e expansion as before to get $\psi (m_i^{\widehat{{\bf %
a}}};m_n^{\widehat{{\bf c}}})$, we then obtain the following probability
amplitudes

\begin{equation}
\psi ((+\frac 12)^{(\widehat{{\bf a}})};(+\frac 12)^{(\widehat{{\bf c}}%
)})=e^{i(\varphi ^{\prime }-\varphi ^{\prime \prime })}\cos \theta ^{\prime
\prime }/2\cos \theta ^{\prime }/2+\sin \theta ^{\prime \prime }/2\sin
\theta ^{\prime }/2  \label{fo42b}
\end{equation}

\begin{equation}
\psi ((+\frac 12)^{(\widehat{{\bf a}})};(-\frac 12)^{(\widehat{{\bf c}}%
)})=e^{-i\varphi ^{\prime \prime }}\cos \theta ^{\prime \prime }/2\sin
\theta ^{\prime }/2-e^{-i\varphi ^{\prime }}\sin \theta ^{\prime \prime
}/2\cos \theta ^{\prime }/2  \label{fo42c}
\end{equation}

\begin{equation}
\psi ((-\frac 12)^{(\widehat{{\bf a}})};(+\frac 12)^{(\widehat{{\bf c}}%
)})=e^{i\varphi ^{\prime }}\cos \theta ^{\prime }/2\sin \theta ^{\prime
\prime }/2-e^{i\varphi ^{\prime \prime }}\sin \theta ^{\prime }/2\cos \theta
^{\prime \prime }/2  \label{fo43}
\end{equation}
and

\begin{equation}
\psi ((-\frac 12)^{(\widehat{{\bf a}})};(-\frac 12)^{(\widehat{{\bf c}}%
)})=e^{-i(\varphi ^{\prime }-\varphi ^{\prime \prime })}\cos \theta ^{\prime
\prime }/2\cos \theta ^{\prime }/2+\sin \theta ^{\prime \prime }/2\sin
\theta ^{\prime }/2  \label{fo44}
\end{equation}

These probability amplitudes of course lead to the same probabilities as
Eqns. (\ref{ni19}) - (\ref{tw22}), namely $[1]$

\begin{eqnarray}
\left| \psi ((+\frac 12)^{(\widehat{{\bf a}})};(+\frac 12)^{(\widehat{{\bf c}%
})})\right| ^2 &=&\left| \psi ((-\frac 12)^{(\widehat{{\bf a}})};(-\frac
12)^{(\widehat{{\bf c}})})\right| ^2  \nonumber \\
&=&\cos ^2(\theta ^{\prime \prime }-\theta ^{\prime })/2-\sin \theta
^{\prime \prime }\sin \theta ^{\prime }\sin ^2(\varphi ^{\prime }-\varphi
^{\prime \prime })/2  \label{fo45}
\end{eqnarray}
and 
\begin{eqnarray}
\left| \psi ((+\frac 12)^{(\widehat{{\bf a}})};(-\frac 12)^{(\widehat{{\bf c}%
})})\right| ^2 &=&\left| \psi ((-\frac 12)^{(\widehat{{\bf a}})};(+\frac
12)^{(\widehat{{\bf c}})})\right| ^2  \nonumber \\
&=&\sin ^2(\theta ^{\prime \prime }-\theta ^{\prime })/2+\sin \theta
^{\prime \prime }\sin \theta ^{\prime }\sin ^2(\varphi ^{\prime }-\varphi
^{\prime \prime })/2  \label{fo46}
\end{eqnarray}

\subsection{New $z$ Component of Spin}

We now plug the new probability amplitudes $\chi (m_i^{\widehat{{\bf b}}%
};m_n^{\widehat{{\bf c}}})$, Eqns. (\ref{fo42b}) - Eqn. (\ref{fo44}), into
the expressions Eqns. (\ref{tw2}) - (\ref{fi5}) for the elements of $[\sigma
_{\widehat{{\bf c}}}]$, the generalized component of $[\sigma _z]$. We find
that

\begin{equation}
(\sigma _{\widehat{{\bf c}}})_{11}=\cos \theta \cos \theta ^{\prime }+\sin
\theta \sin \theta ^{\prime }\cos (\varphi -\varphi ^{\prime })  \label{fo47}
\end{equation}

\begin{equation}
(\sigma _{\widehat{{\bf c}}})_{12}=\sin \theta \cos \theta ^{\prime
}e^{i\varphi }+\sin \theta ^{\prime }\sin ^2\tfrac \theta 2e^{i\varphi
^{\prime }}-\sin \theta ^{\prime }\cos ^2\tfrac \theta 2e^{i(2\varphi
-\varphi ^{\prime })}  \label{fo48}
\end{equation}

\begin{equation}
(\sigma _{\widehat{{\bf c}}})_{21}=\sin \theta \cos \theta ^{\prime
}e^{-i\varphi }+\sin \theta ^{\prime }\sin ^2\tfrac \theta 2e^{-i\varphi
^{\prime }}-\sin \theta ^{\prime }\cos ^2\tfrac \theta 2e^{-i(2\varphi
-\varphi ^{\prime })}  \label{fo49}
\end{equation}

\begin{equation}
(\sigma _{\widehat{{\bf c}}})_{22}=-\cos \theta \cos \theta ^{\prime }-\sin
\theta \sin \theta ^{\prime }\cos (\varphi -\varphi ^{\prime })  \label{fi50}
\end{equation}

According to the reasoning given in Ref. $1$, the eigenvectors of this
operator have for their elements generalized probability amplitudes. Thus
the form of these elements is given by Eqns. (\ref{fo42b}) - (\ref{fo44}).
However, the arguments have to changeso that $\widehat{{\bf a}}\rightarrow 
\widehat{{\bf c}}$ and $\widehat{{\bf c}}\rightarrow \widehat{{\bf b}}$.
Hence the eigenvectors are

\begin{equation}
\lbrack \xi _{\widehat{{\bf c}}}^{(+)}]=\left( 
\begin{array}{c}
\chi ((+\frac 12)^{(\widehat{{\bf c}})};(+\frac 12)^{(\widehat{{\bf b}})})
\\ 
\chi ((+\frac 12)^{(\widehat{{\bf c}})};(-\frac 12)^{(\widehat{{\bf b}})})
\end{array}
\right) =\left( 
\begin{array}{c}
\cos \frac \theta 2\cos \frac{\theta ^{\prime }}2e^{i(\varphi -\varphi
^{\prime })}+\sin \frac \theta 2\sin \frac{\theta ^{\prime }}2 \\ 
\sin \frac \theta 2\cos \frac{\theta ^{\prime }}2e^{-i\varphi ^{\prime
}}-\cos \frac \theta 2\sin \frac{\theta ^{\prime }}2e^{-i\varphi }
\end{array}
\right)  \label{fi51}
\end{equation}
for eigenvalue $+1$, and 
\begin{equation}
\lbrack \xi _{\widehat{{\bf c}}}^{(-)}]=\left( 
\begin{array}{c}
\chi ((-\frac 12)^{(\widehat{{\bf c}})};(+\frac 12)^{(\widehat{{\bf b}})})
\\ 
\chi ((-\frac 12)^{(\widehat{{\bf c}})};(-\frac 12)^{(\widehat{{\bf b}})})
\end{array}
\right) =\left( 
\begin{array}{c}
\sin \frac{\theta ^{\prime }}2\cos \frac \theta 2e^{i\varphi }-\cos \frac{%
\theta ^{\prime }}2\sin \frac \theta 2e^{i\varphi ^{\prime }} \\ 
\cos \frac \theta 2\cos \frac{\theta ^{\prime }}2e^{-i(\varphi -\varphi
^{\prime })}+\sin \frac \theta 2\sin \frac{\theta ^{\prime }}2
\end{array}
\right)  \label{fi52}
\end{equation}
for eigenvalue $-1$. Direct calculation confirms that, indeed,

\begin{equation}
\lbrack \sigma _z][\xi _{\widehat{{\bf c}}}^{(\pm )}]=\pm [\xi _{\widehat{%
{\bf c}}}^{(\pm )}].  \label{fi53}
\end{equation}

\subsection{New $x$ and $y$ Components of Spin}

As demonstrated in Refs. $2$ and $3$, the $x$ component of spin and its
eigenvectors are obtainable from the $z$ component and its eigenvectors
through the transformation $\theta ^{\prime }\rightarrow \theta ^{\prime
}-\pi /2$ for the old choice of phase $.$ The $y$ component and its
eigenvectors are obtainable through the transformations $\theta ^{\prime
}=\pi /2$, $\varphi ^{\prime }\rightarrow \varphi ^{\prime }-\pi /2.$ But
these transformations do not apply to the new choice of phase. When we use
them, the resulting operators together with the $z$ component do not satisfy
the commutation relations. We therefore resort to the more laborious but
sure method of obtaining the $x$ and $y$ components through the ladder
operators. Using the properties 
\begin{equation}
\lbrack \sigma _{+}][\xi _{\widehat{{\bf c}}}^{(+)}]=0;\;\;\;[\sigma
_{+}][\xi _{\widehat{{\bf c}}}^{(-)}]=2[\xi _{\widehat{{\bf c}}}^{(+)}]\;
\label{fi54}
\end{equation}
and 
\begin{equation}
\lbrack \sigma _{-}][\xi _{\widehat{{\bf c}}}^{(-)}]=0;\;\;\;[\sigma
_{-}][\xi _{\widehat{{\bf c}}}^{(+)}]=2[\xi _{\widehat{{\bf c}}}^{(-)}],\;
\label{fi55}
\end{equation}
we find that the elements of $[\sigma _{+}]$ are

\begin{equation}
(\sigma _{+})_{11}=\cos \theta \sin \theta ^{\prime }e^{-i\varphi ^{\prime
}}+\sin \theta \sin ^2\tfrac{\theta ^{\prime }}2e^{-i\varphi }-\sin \theta
\cos ^2\tfrac{\theta ^{\prime }}2e^{i(\varphi -2\varphi ^{\prime })}
\label{fi56}
\end{equation}
\begin{equation}
(\sigma _{+})_{12}=2e^{2i(\varphi -\varphi ^{\prime })}\cos ^2\tfrac \theta
2\cos ^2\tfrac{\theta ^{\prime }}2+\sin \theta \sin \theta ^{\prime
}e^{i(\varphi -\varphi ^{\prime })}+2\sin ^2\tfrac \theta 2\sin ^2\tfrac{%
\theta ^{\prime }}2  \label{fi57}
\end{equation}
\begin{equation}
(\sigma _{+})_{21}=-2e^{-2i\varphi ^{\prime }}\sin ^2\tfrac \theta 2\cos ^2%
\tfrac{\theta ^{\prime }}2-2e^{-2i\varphi }\cos ^2\tfrac \theta 2\sin ^2%
\tfrac{\theta ^{\prime }}2+\sin \theta \sin \theta ^{\prime }e^{-i(\varphi
+\varphi ^{\prime })}  \label{fi58}
\end{equation}
and

\begin{equation}
(\sigma _{+})_{22}=-\cos \theta \sin \theta ^{\prime }e^{-i\varphi ^{\prime
}}-\sin \theta \sin ^2\tfrac{\theta ^{\prime }}2e^{-i\varphi }+\sin \theta
\cos ^2\tfrac{\theta ^{\prime }}2e^{i(\varphi -2\varphi ^{\prime })}.
\label{fi59}
\end{equation}

The elements of $[\sigma _{-}]$ are

\begin{equation}
(\sigma _{-})_{11}=\sin \theta \sin ^2\tfrac{\theta ^{\prime }}2e^{i\varphi
}+\sin \theta ^{\prime }\cos e^{i\varphi ^{\prime }}-\sin \theta \cos ^2%
\tfrac{\theta ^{\prime }}2e^{i(2\varphi ^{\prime }-\varphi )}  \label{si60}
\end{equation}

\begin{equation}
(\sigma _{-})_{12}=-2e^{2i\varphi }\cos ^2\tfrac \theta 2\sin ^2\tfrac{%
\theta ^{\prime }}2-2e^{2i\varphi ^{\prime }}\sin ^2\tfrac \theta 2\cos ^2%
\tfrac{\theta ^{\prime }}2+\sin \theta \sin \theta ^{\prime }e^{i(\varphi
+\varphi ^{\prime })}  \label{si61}
\end{equation}

\begin{equation}
(\sigma _{-})_{21}=2e^{2i(\varphi ^{\prime }-\varphi )}\cos ^2\tfrac \theta
2\cos ^2\tfrac{\theta ^{\prime }}2+\sin \theta \sin \theta ^{\prime
}e^{i(\varphi ^{\prime }-\varphi )}+2\sin ^2\tfrac \theta 2\sin ^2\tfrac{%
\theta ^{\prime }}2  \label{si62}
\end{equation}
and 
\begin{equation}
(\sigma _{-})_{22}=-\sin \theta \sin ^2\tfrac{\theta ^{\prime }}2e^{i\varphi
}-\sin \theta ^{\prime }\cos e^{i\varphi ^{\prime }}+\sin \theta \cos ^2%
\tfrac{\theta ^{\prime }}2e^{i(2\varphi ^{\prime }-\varphi )}.  \label{si63}
\end{equation}

Hence, the elements of $[\sigma _x]$, obtained using 
\begin{equation}
\lbrack \sigma _x]=\frac 12([\sigma _{+}]+[\sigma _{-}]),  \label{si64}
\end{equation}
are

\begin{eqnarray}
(\sigma _x)_{11} &=&\frac 12\sin \theta ^{\prime }\cos \theta e^{i\varphi
^{\prime }}+\frac 12\sin \theta ^{\prime }\cos \theta e^{-i\varphi ^{\prime
}}+\frac 12\sin \theta \sin ^2\tfrac{\theta ^{\prime }}2e^{i\varphi } 
\nonumber \\
&&+\frac 12\sin \theta \sin ^2\tfrac{\theta ^{\prime }}2e^{-i\varphi }-\frac
12\sin \theta \cos ^2\tfrac{\theta ^{\prime }}2e^{i(2\varphi ^{\prime
}-\varphi )}  \nonumber \\
&&-\frac 12\sin \theta \cos ^2\tfrac{\theta ^{\prime }}2e^{i(\varphi
-2\varphi ^{\prime })}  \label{si65}
\end{eqnarray}

\begin{eqnarray}
(\sigma _x)_{12} &=&\cos ^2\tfrac \theta 2\cos ^2\tfrac{\theta ^{\prime }}%
2e^{2i(\varphi -\varphi ^{\prime })}+\frac 12\sin \theta \sin \theta
^{\prime }e^{i(\varphi -\varphi ^{\prime })}-\cos ^2\tfrac \theta 2\sin ^2%
\tfrac{\theta ^{\prime }}2e^{2i\varphi }  \nonumber \\
&&-\sin ^2\tfrac \theta 2\cos ^2\tfrac{\theta ^{\prime }}2e^{2i\varphi
^{\prime }}+\frac 12\sin \theta \sin \theta ^{\prime }e^{i(\varphi +\varphi
^{\prime })}+\sin ^2\tfrac \theta 2\sin ^2\tfrac{\theta ^{\prime }}2
\label{si66}
\end{eqnarray}

\begin{eqnarray}
(\sigma _x)_{21} &=&\cos ^2\tfrac \theta 2\cos ^2\tfrac{\theta ^{\prime }}%
2e^{2i(\varphi ^{\prime }-\varphi )}+\frac 12\sin \theta \sin \theta
^{\prime }e^{i(\varphi ^{\prime }-\varphi )}-\cos ^2\tfrac \theta 2\sin ^2%
\tfrac{\theta ^{\prime }}2e^{-2i\varphi }  \nonumber \\
&&\ -\sin ^2\tfrac \theta 2\cos ^2\tfrac{\theta ^{\prime }}2e^{-2i\varphi
^{\prime }}+\frac 12\sin \theta \sin \theta ^{\prime }e^{-i(\varphi +\varphi
^{\prime })}+\sin ^2\tfrac \theta 2\sin ^2\tfrac{\theta ^{\prime }}2
\label{si67}
\end{eqnarray}
and 
\begin{eqnarray}
(\sigma _x)_{22} &=&-\frac 12\sin \theta ^{\prime }\cos \theta e^{i\varphi
^{\prime }}-\frac 12\sin \theta ^{\prime }\cos \theta e^{-i\varphi ^{\prime
}}-\frac 12\sin \theta \sin ^2\tfrac{\theta ^{\prime }}2e^{i\varphi } 
\nonumber \\
&&\ -\frac 12\sin \theta \sin ^2\tfrac{\theta ^{\prime }}2e^{-i\varphi
}+\frac 12\sin \theta \cos ^2\tfrac{\theta ^{\prime }}2e^{i(2\varphi
^{\prime }-\varphi )}  \nonumber \\
&&+\frac 12\sin \theta \cos ^2\tfrac{\theta ^{\prime }}2e^{i(\varphi
-2\varphi ^{\prime })}  \label{si68}
\end{eqnarray}

The elements of $[\sigma _y]$, obtained from

\begin{equation}
\lbrack \sigma _y]=\frac 1{2i}([\sigma _{+}]-[\sigma _{-}]),  \label{si69}
\end{equation}
are 
\begin{eqnarray}
(\sigma _y)_{11} &=&\frac i2[\sin \theta ^{\prime }\cos \theta e^{i\varphi
^{\prime }}-\sin \theta ^{\prime }\cos \theta e^{-i\varphi ^{\prime }}+\sin
\theta \sin ^2\tfrac{\theta ^{\prime }}2e^{i\varphi }  \nonumber \\
&&\ \ -\sin \theta \sin ^2\tfrac{\theta ^{\prime }}2e^{-i\varphi }+\sin
\theta \cos ^2\tfrac{\theta ^{\prime }}2e^{i(\varphi -2\varphi ^{\prime })} 
\nonumber \\
&&\ -\sin \theta \cos ^2\tfrac{\theta ^{\prime }}2e^{i(2\varphi ^{\prime
}-\varphi )}]  \label{se70}
\end{eqnarray}

\begin{eqnarray}
(\sigma _y)_{12} &=&-i[\cos ^2\tfrac \theta 2\cos ^2\tfrac{\theta ^{\prime }}%
2e^{2i(\varphi -\varphi ^{\prime })}+\sin ^2\tfrac \theta 2\sin ^2\tfrac{%
\theta ^{\prime }}2+\frac 12\sin \theta \sin \theta ^{\prime }e^{i(\varphi
-\varphi ^{\prime })}  \nonumber \\
&&+\cos ^2\tfrac \theta 2\sin ^2\tfrac{\theta ^{\prime }}2e^{2i\varphi
}+\sin ^2\tfrac \theta 2\cos ^2\tfrac{\theta ^{\prime }}2e^{2i\varphi
^{\prime }}  \nonumber \\
&&-\frac 12\sin \theta \sin \theta ^{\prime }e^{i(\varphi +\varphi ^{\prime
})}]  \label{se71}
\end{eqnarray}

\begin{eqnarray}
(\sigma _y)_{21} &=&i[\cos ^2\tfrac \theta 2\cos ^2\tfrac{\theta ^{\prime }}%
2e^{2i(\varphi ^{\prime }-\varphi )}+\sin ^2\tfrac \theta 2\sin ^2\tfrac{%
\theta ^{\prime }}2+\frac 12\sin \theta \sin \theta ^{\prime }e^{i(\varphi
^{\prime }-\varphi )}  \nonumber \\
&&+\cos ^2\tfrac \theta 2\sin ^2\tfrac{\theta ^{\prime }}2e^{-2i\varphi
}+\sin ^2\tfrac \theta 2\cos ^2\tfrac{\theta ^{\prime }}2e^{-2i\varphi
^{\prime }}  \nonumber \\
&&-\frac 12\sin \theta \sin \theta ^{\prime }e^{-i(\varphi +\varphi ^{\prime
})}]  \label{se72}
\end{eqnarray}
and 
\begin{eqnarray}
(\sigma _y)_{22} &=&-\frac i2[\sin \theta ^{\prime }\cos \theta e^{i\varphi
^{\prime }}-\sin \theta ^{\prime }\cos \theta e^{-i\varphi ^{\prime }}+\sin
\theta \sin ^2\tfrac{\theta ^{\prime }}2e^{i\varphi }  \nonumber \\
&&-\sin \theta \sin ^2\tfrac{\theta ^{\prime }}2e^{-i\varphi }+\sin \theta
\cos ^2\tfrac{\theta ^{\prime }}2e^{i(\varphi -2\varphi ^{\prime })} 
\nonumber \\
&&-\sin \theta \cos ^2\tfrac{\theta ^{\prime }}2e^{i(2\varphi ^{\prime
}-\varphi )}]  \label{se73}
\end{eqnarray}

The three spin operators possess the usual properties. Thus each one gives
the unit matrix when squared.

\begin{equation}
\lbrack \sigma _x]^2=[\sigma _y]^2=[\sigma _{\widehat{{\bf c}}}]^2=I
\label{se74}
\end{equation}
The commutators of these matrices are

\begin{equation}
\lbrack [\sigma _i],[\sigma _j]]=2i[\sigma _k],  \label{se75}
\end{equation}
where $i,j,k=x,y,z$ are taken in cyclic permutation. Finally, the operators
anti-commute:

\begin{equation}
\lbrack [\sigma _i],[\sigma _j]]_{+}=0  \label{se76}
\end{equation}
for $i\neq j.$

\subsection{Eigenvectors of the $x$ and $y$ Components of Spin}

The surest way of obtaining the eigenvectors of $[\sigma _x]$ and $[\sigma
_y]$ is by means of rotations. Thus, if the $z$ axis is rotated through the
angle $\pi /2$ in the positive sense about the $y$ axis, it becomes the $x$
axis. Its rotated eigenvectors become the eigenvectors of $[\sigma _x].$ The
eigenvectors are obtained through the formula [$8$]

\begin{equation}
\lbrack \xi _x^{(\pm )}]=\frac 1{\sqrt{2}}(I-i[\sigma _y])[\xi _{\widehat{%
{\bf c}}}^{(\pm )}],  \label{se77}
\end{equation}
where $I$ is the unit $2\times 2$ matrix.Using Eqns. (\ref{se70}) - (\ref
{se73}) for the elements of $[\sigma _y]$, we find that the eigenvectors of $%
[\sigma _x]$ are 
\begin{equation}
\lbrack \xi _x^{(+)}]=\frac 1{\sqrt{2}}\left( 
\begin{array}{c}
\cos \frac \theta 2\cos \frac{\theta ^{\prime }}2e^{i(\varphi -\varphi
^{\prime })}+\sin \frac{\theta ^{\prime }}2\cos \frac \theta 2e^{i\varphi
}-\sin \frac \theta 2\cos \frac{\theta ^{\prime }}2e^{i\varphi ^{\prime
}}+\sin \frac \theta 2\sin \frac{\theta ^{\prime }}2 \\ 
\cos \frac \theta 2\cos \frac{\theta ^{\prime }}2e^{i(\varphi ^{\prime
}-\varphi )}-\sin \frac{\theta ^{\prime }}2\cos \frac \theta 2e^{-i\varphi
}+\sin \frac \theta 2\cos \frac{\theta ^{\prime }}2e^{-i\varphi ^{\prime
}}+\sin \frac \theta 2\sin \frac{\theta ^{\prime }}2
\end{array}
\right)  \label{se78}
\end{equation}
for eigenvalue $+1$ and

\begin{equation}
\lbrack \xi _x^{(-)}]=\frac 1{\sqrt{2}}\left( 
\begin{array}{c}
-\cos \frac \theta 2\cos \frac{\theta ^{\prime }}2e^{i(\varphi -\varphi
^{\prime })}+\sin \frac{\theta ^{\prime }}2\cos \frac \theta 2e^{i\varphi
}-\sin \frac \theta 2\cos \frac{\theta ^{\prime }}2e^{i\varphi ^{\prime
}}-\sin \frac \theta 2\sin \frac{\theta ^{\prime }}2 \\ 
\cos \frac \theta 2\cos \frac{\theta ^{\prime }}2e^{i(\varphi ^{\prime
}-\varphi )}+\sin \frac{\theta ^{\prime }}2\cos \frac \theta 2e^{-i\varphi
}-\sin \frac \theta 2\cos \frac{\theta ^{\prime }}2e^{-i\varphi ^{\prime
}}+\sin \frac \theta 2\sin \frac{\theta ^{\prime }}2
\end{array}
\right)  \label{se79}
\end{equation}
for eigenvalue $-1$.

For the $z$ axis to change into the $y$ axis, we need to rotate it in the
negative sense about the $x$ axis through an angle of $\pi /2$. Thus the
eigenvectors of $[\sigma _y]$ are given by

\begin{equation}
\lbrack \xi _y^{(\pm )}]=\frac 1{\sqrt{2}}(I+i[\sigma _x])[\xi _{\widehat{%
{\bf c}}}^{(\pm )}].  \label{ei80}
\end{equation}
Hence, using Eqns. (\ref{si65}) - (\ref{si68}) for the elements of $[\sigma
_x],$ we find that the eigenvectors are

\begin{equation}
\lbrack \xi _y^{(+)}]=\left( 
\begin{array}{c}
\cos \frac \theta 2\cos \frac{\theta ^{\prime }}2e^{i(\varphi -\varphi
^{\prime })}+i\sin \frac{\theta ^{\prime }}2\cos \frac \theta 2e^{i\varphi
}-i\sin \frac \theta 2\cos \frac{\theta ^{\prime }}2e^{i\varphi ^{\prime
}}+\sin \frac \theta 2\sin \frac{\theta ^{\prime }}2 \\ 
i\cos \frac \theta 2\cos \frac{\theta ^{\prime }}2e^{i(\varphi ^{\prime
}-\varphi )}-\sin \frac{\theta ^{\prime }}2\cos \frac \theta 2e^{-i\varphi
}+\sin \frac \theta 2\cos \frac{\theta ^{\prime }}2e^{-i\varphi ^{\prime
}}+i\sin \frac \theta 2\sin \frac{\theta ^{\prime }}2
\end{array}
\right)  \label{ei81}
\end{equation}
for the eigenvalue $+1$, and 
\begin{equation}
\lbrack \xi _y^{(-)}]=\left( 
\begin{array}{c}
i\cos \frac \theta 2\cos \frac{\theta ^{\prime }}2e^{i(\varphi -\varphi
^{\prime })}+\sin \frac{\theta ^{\prime }}2\cos \frac \theta 2e^{i\varphi
}-\sin \frac \theta 2\cos \frac{\theta ^{\prime }}2e^{i\varphi ^{\prime
}}+i\sin \frac \theta 2\sin \frac{\theta ^{\prime }}2 \\ 
\cos \frac \theta 2\cos \frac{\theta ^{\prime }}2e^{i(\varphi ^{\prime
}-\varphi )}-i\sin \frac{\theta ^{\prime }}2\cos \frac \theta 2e^{-i\varphi
}+i\sin \frac \theta 2\cos \frac{\theta ^{\prime }}2e^{-i\varphi ^{\prime
}}+\sin \frac \theta 2\sin \frac{\theta ^{\prime }}2
\end{array}
\right)  \label{ei82}
\end{equation}
for the eigenvalue $-1$.

\section{Reduction to Standard Forms}

Since we know the standard forms of the spin operators and their
eigenvectors, we can deduce the limit in which the generalized operators and
their eigenvectors reduce to these standard forms. To obtain the Pauli
matrices and their eigenvectors, we find that we need to set $\theta =\theta
^{\prime }$ and $\varphi =\varphi ^{\prime }$. Thus, we find that in this
limit,

\begin{equation}
\lbrack \sigma _{\widehat{{\bf c}}}]\rightarrow [\sigma _z]=\left( 
\begin{array}{cc}
1 & 0 \\ 
0 & -1
\end{array}
\right) ,  \label{ei83}
\end{equation}
with the eigenvectors

\begin{equation}
\lbrack \xi _{\widehat{{\bf c}}}^{(+)}]\rightarrow \left( 
\begin{array}{c}
1 \\ 
0
\end{array}
\right) \text{ \ \ \ \ and \ \ \ }[\xi _{\widehat{{\bf c}}}^{(-)}%
]\rightarrow \left( 
\begin{array}{c}
0 \\ 
1
\end{array}
\right) \text{\ \ }  \label{ei84}
\end{equation}
for the eigenvalues $+1$ and $-1$ respectively. The $x$ component becomes 
\begin{equation}
\lbrack \sigma _x]=\left( 
\begin{array}{cc}
0 & 1 \\ 
1 & 0
\end{array}
\right)  \label{ei86}
\end{equation}
with the eigenvectors 
\begin{equation}
\lbrack \xi _x^{(+)}]=\frac 1{\sqrt{2}}\left( 
\begin{array}{c}
1 \\ 
1
\end{array}
\right) \text{ \ \ \ \ and \ \ \ }[\xi _x^{(-)}]=\frac 1{\sqrt{2}}\left( 
\begin{array}{c}
-1 \\ 
1
\end{array}
\right) \text{\ \ }  \label{ei84a}
\end{equation}
for the eigenvalues $+1$ and $-1$ respectively. The $y$ component becomes 
\begin{equation}
\lbrack \sigma _y]=\left( 
\begin{array}{cc}
0 & -i \\ 
i & 0
\end{array}
\right) ,  \label{ei87}
\end{equation}
with the eigenvectors

\begin{equation}
\lbrack \xi _y^{(+)}]=\frac 1{\sqrt{2}}\left( 
\begin{array}{c}
1 \\ 
i
\end{array}
\right) \text{ \ \ \ \ and \ \ \ }[\xi _y^{(-)}]=\frac 1{\sqrt{2}}\left( 
\begin{array}{c}
i \\ 
1
\end{array}
\right) \text{\ \ }  \label{ei84b}
\end{equation}
for the eigenvalues $+1$ and $-1$ respectively.

To obtain the ''standard generalized form'' for $[\sigma _{\widehat{{\bf c}}%
}]$, Eqn. (\ref{tw12}), we find that we have to set $\theta =0,$ $\varphi
=\frac \pi 2$. We then recover Eqn. (\ref{tw12}); its eigenvectors for the
respective eigenvalues $+1$ and $-1$ are found to be

\begin{equation}
\lbrack \xi _{\widehat{{\bf c}}}^{(+)}]=i\left( 
\begin{array}{c}
\cos \frac{\theta ^{\prime }}2e^{-i\varphi ^{\prime }} \\ 
\sin \frac{\theta ^{\prime }}2
\end{array}
\right)  \label{ei88}
\end{equation}
and 
\begin{equation}
\lbrack \xi _{\widehat{{\bf c}}}^{(-)}]=i\left( 
\begin{array}{c}
\sin \frac{\theta ^{\prime }}2 \\ 
-\cos \frac{\theta ^{\prime }}2e^{-i\varphi ^{\prime }}
\end{array}
\right) ,  \label{ei89}
\end{equation}
which differ from Eqns. (\ref{fo41z}) and (\ref{fo42}) through the
appearance of the phase factor $i$ (apart from the change of argument
arising from the change in quantization direction from $\widehat{{\bf a}}$
to $\widehat{{\bf c}}$).

In the same limit, we obtain 
\begin{equation}
\lbrack \sigma _x]=\left( 
\begin{array}{cc}
\sin \theta ^{\prime }\cos \varphi ^{\prime } & \sin ^2\frac{\theta ^{\prime
}}2-\cos ^2\frac{\theta ^{\prime }}2e^{-2i\varphi ^{\prime }} \\ 
\sin ^2\frac{\theta ^{\prime }}2-\cos ^2\frac{\theta ^{\prime }}%
2e^{2i\varphi ^{\prime }} & -\sin \theta ^{\prime }\cos \varphi ^{\prime }
\end{array}
\right)  \label{ni90}
\end{equation}
with the eigenvectors

\begin{equation}
\lbrack \xi _x^{(+)}]=\frac i{\sqrt{2}}\left( 
\begin{array}{c}
\sin \frac{\theta ^{\prime }}2+\cos \frac{\theta ^{\prime }}2e^{-\varphi
^{\prime }} \\ 
\sin \frac{\theta ^{\prime }}2-\cos \frac{\theta ^{\prime }}2e^{\varphi
^{\prime }}
\end{array}
\right)  \label{ni91}
\end{equation}
and 
\begin{equation}
\lbrack \xi _x^{(-)}]=-\frac i{\sqrt{2}}\left( 
\begin{array}{c}
\cos \frac{\theta ^{\prime }}2e^{-\varphi ^{\prime }}-\sin \frac{\theta
^{\prime }}2 \\ 
\cos \frac{\theta ^{\prime }}2e^{\varphi ^{\prime }}+\sin \frac{\theta
^{\prime }}2
\end{array}
\right) .  \label{ni92}
\end{equation}

We also obtain in this limit 
\begin{equation}
\lbrack \sigma _y]=\left( 
\begin{array}{cc}
-\sin \theta ^{\prime }\cos \varphi ^{\prime } & i[\cos ^2\frac{\theta
^{\prime }}2e^{-2i\varphi ^{\prime }}+\sin ^2\frac{\theta ^{\prime }}2] \\ 
-i[\cos ^2\frac{\theta ^{\prime }}2e^{2i\varphi ^{\prime }}+\sin ^2\frac{%
\theta ^{\prime }}2] & \sin \theta ^{\prime }\cos \varphi ^{\prime }
\end{array}
\right)  \label{ni93}
\end{equation}
with the eigenvectors 
\begin{equation}
\lbrack \xi _y^{(+)}]=\frac 1{\sqrt{2}}\left( 
\begin{array}{c}
i\cos \frac{\theta ^{\prime }}2e^{-\varphi ^{\prime }}-\sin \frac{\theta
^{\prime }}2 \\ 
\cos \frac{\theta ^{\prime }}2e^{\varphi ^{\prime }}+i\sin \frac{\theta
^{\prime }}2
\end{array}
\right)  \label{ni94}
\end{equation}
and 
\begin{equation}
\lbrack \xi _y^{(-)}]=-\frac 1{\sqrt{2}}\left( 
\begin{array}{c}
-i\sin \frac{\theta ^{\prime }}2+\cos \frac{\theta ^{\prime }}2e^{-\varphi
^{\prime }} \\ 
\sin \frac{\theta ^{\prime }}2+i\cos \frac{\theta ^{\prime }}2e^{\varphi
^{\prime }}
\end{array}
\right) .  \label{ni95}
\end{equation}

We observe that in order to recover the Pauli matrices and their
eigenvectors from these ''generalized standard forms'', we have to set $%
\theta ^{\prime }=0$ and $\varphi ^{\prime }=\pi /2$, so as to achieve the
conditions $\theta =\theta ^{\prime }$ and $\varphi =\varphi ^{\prime }.$

\section{Discussion and Conclusion}

In this paper, we have derived new forms for the operators and states
describing spin $1/2$. Thus,we now have two sets of forms for the
generalized probability amplitudes and correspondingly for the operators.
Since there are other choices of phase for the eigenvectors of Eqn. (\ref
{tw12}), it follows that there are yet more forms of the probability
amplitudes, operators and their eigenvectors. Using the methods illustrated
here and in Refs. $1-3$, it is straightforward, but perhaps tedious, to
obtain the spin quantities for any choice of phase. It is probable that in
some cases it will be possible to obtain the $x$ and $y$ components of the
operators from the $z$ components by straightforward substitution of angles,
as we did for the original choice of phase [$2,3$]. It should be mentioned
that this choice was purely arbitrary, and it seems a fortunate accident
that this choice permitted of obtaining the $x$ and $y$ components of spin
from the $z$ component by the procedure of changing the arguments. For any
choice of phase, the method employing the ladder operators will always yield
the $x$ and $y$ components of spin. Since the eigenvectors of $[\sigma _z]$
are easily obtained for any choice of phase, it will always be possible by
using rotations to deduce the eigenvectors of these operators.

Though we have illustrated our considerations using the case of spin $1/2$,
the general theory will work perfectly well for any value of spin $[3]$. Of
course, the total number of phase combinations goes up with value of $J$,
and since in the general case, the $x$ and $y$ components of spin have to be
obtained via the ladder operators, this will result in a great deal of
calculational labour.

It would seem obvious that in any calculations involving these generalized
quantities, it is necessary to match the particular probability amplitudes
with the correct forms of the operators. The correct forms of the operators
are those which result from using the same probability amplitudes when the
elements of the operator are being derived. It is not clear at this time
what the effect would be of mixing one set of probability amplitudes with
operators corresponding to another set.

At this point, it seems accurate to say that whatever form is obtained for
the generalized quantities, the Pauli forms are obtained by setting $\theta
=\theta ^{\prime }$ and $\varphi =\varphi ^{\prime }$. On the other hand,
there does not appear to be a uniform prescription for obtaining the
''standard generalized quantities''; while we must set $\theta $ equal to
zero, the required value of $\varphi $ seems to depend on the exact choice
of phase.

\section{References}

1. Mweene H. V., ''Derivation of Spin Vectors and Operators From First
Principles'', submitted to {\it Foundations of Physics}, quant-ph/9905012

2. Mweene H. V., ''Generalized Spin-1/2 Operators and Their Eigenvectors'',
quant-ph/9906002

3. Mweene H. V., ''Vectors and Operators For Spin 1 Derived From First
Principles'', quant-ph/9906043

4. Land\'e A., ''New Foundations of Quantum Mechanics'', Cambridge
University Press, 1965.

5. Land\'e A., ''From Dualism To Unity in Quantum Physics'', Cambridge
University Press, 1960.

6. Land\'e A., ''Foundations of Quantum Theory'', Yale University Press,
1955.

7. Land\'e A., ''Quantum Mechanics in a New Key'', Exposition Press, 1973.

8. See, for example, Bransden and Joachain, ''Introduction to Quantum
Mechanics'', Longman Scientific \& Technical, 1989.

\end{document}